\documentclass[conference]{IEEEtran}
\IEEEoverridecommandlockouts
\usepackage{cite}
\usepackage{amsmath,amssymb,amsfonts}
\usepackage{algorithmic}
\usepackage{graphicx}
\usepackage{textcomp}
\usepackage{xcolor}
\usepackage{siunitx}
\usepackage{verbatim}
\usepackage[version=4]{mhchem}
\usepackage{tikz}
\usepackage{textcomp}
\usepackage{hyperref}
\def\BibTeX{{\rm B\kern-.05em{\sc i\kern-.025em b}\kern-.08em
    T\kern-.1667em\lower.7ex\hbox{E}\kern-.125emX}}

\newcommand\copyrighttext{%
  \footnotesize \textcopyright 2021 IEEE. Personal use of this material is permitted.
  Permission from IEEE must be obtained for all other uses, in any current or future
  media, including reprinting/republishing this material for advertising or promotional
  purposes, creating new collective works, for resale or redistribution to servers or
  lists, or reuse of any copyrighted component of this work in other works.
  }
\newcommand\copyrightnotice{%
\begin{tikzpicture}[remember picture,overlay]
\node[anchor=south,yshift=10pt] at (current page.south) {\fbox{\parbox{\dimexpr\textwidth-\fboxsep-\fboxrule\relax}{\copyrighttext}}};
\end{tikzpicture}%
}    
    
\begin{document}


\title{Inductive Power Transfer Through Saltwater\\
\thanks{We thank NSERC for providing funding for this project.}
}

\author{\IEEEauthorblockN{J.N. Wandinger, D.M. Roberts, J.S. Bobowski}
\IEEEauthorblockA{\textit{Department of Physics} \\
\textit{University of British Columbia}\\
Kelowna, Canada \\
email: jake.bobowski@ubc.ca}
\and
\IEEEauthorblockN{T. Johnson}
\IEEEauthorblockA{\textit{School of Engineering} \\
\textit{University of British Columbia}\\
Kelowna, Canada \\
email: thomas.johnson@ubc.ca}
}

\maketitle
\copyrightnotice

\begin{abstract}
We investigated inductive power transfer (IPT) through a rectangular slab of saltwater.  Our inductively-coupled transmitters and receivers were made from loop-gap resonators (LGRs) having resonant frequencies near 100~MHz.  Electric fields are confined within the narrow gaps of the LGRs making it possible to strongly suppress the power dissipation associated with electric fields in a conductive medium.  Therefore, the power transfer efficiency in our system was limited  by magnetic field dissipation in the conducting medium.  We measured the power transfer efficiency as a function of both the conductivity of the water and the resonant frequency of the LGRs.  We also present an equivalent circuit model that can be used to model IPT through a conductive medium.  Finally, we show that using dividers to partition the saltwater volume provides another means of enhancing power transfer efficiency.   
\end{abstract}


\section{Introduction}
\begin{figure*}[t]
\centerline{
\begin{tabular}{cc}
(a)~\includegraphics[width=0.5\columnwidth]{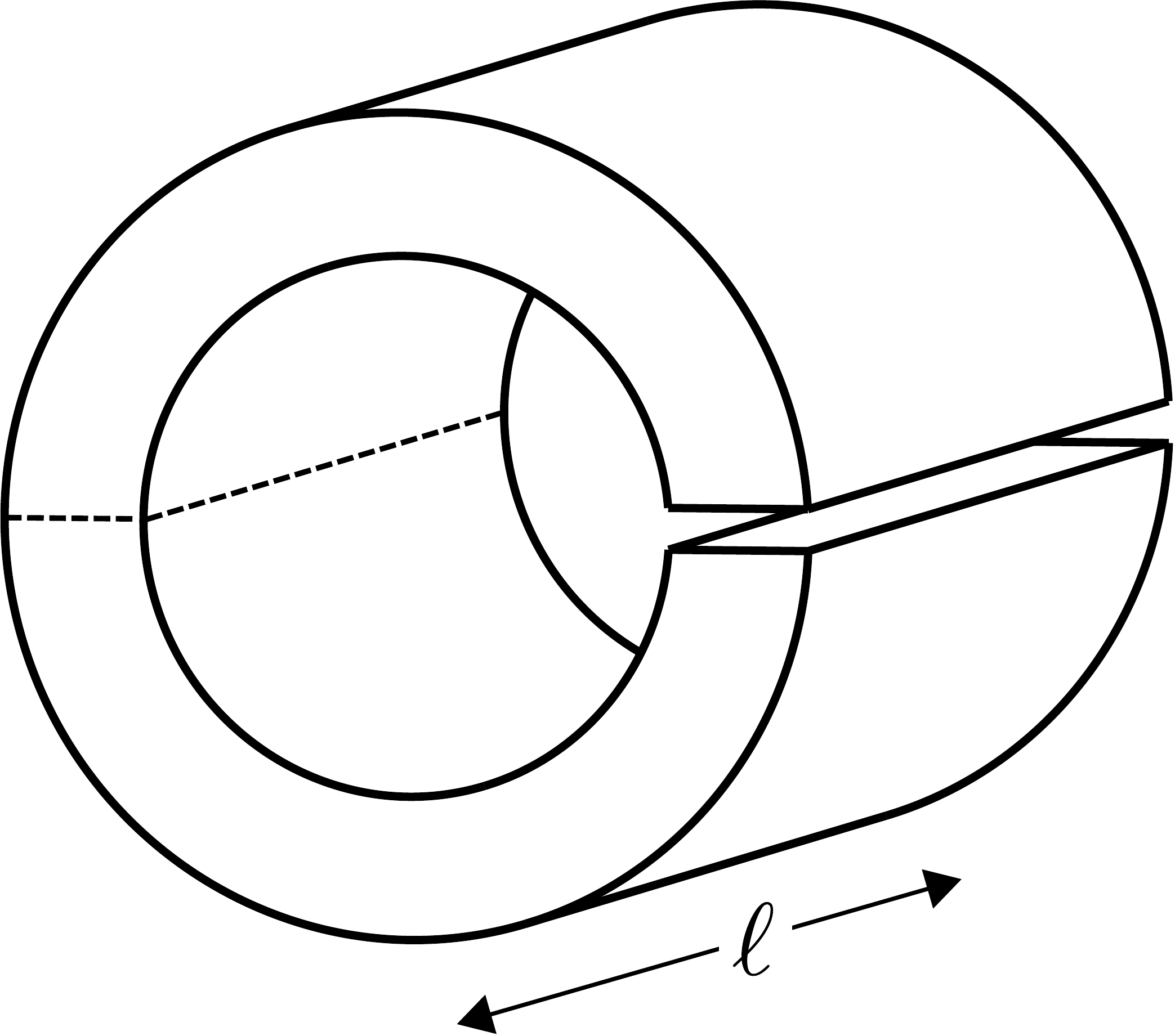} & \qquad(b)~\includegraphics[width=.7\columnwidth]{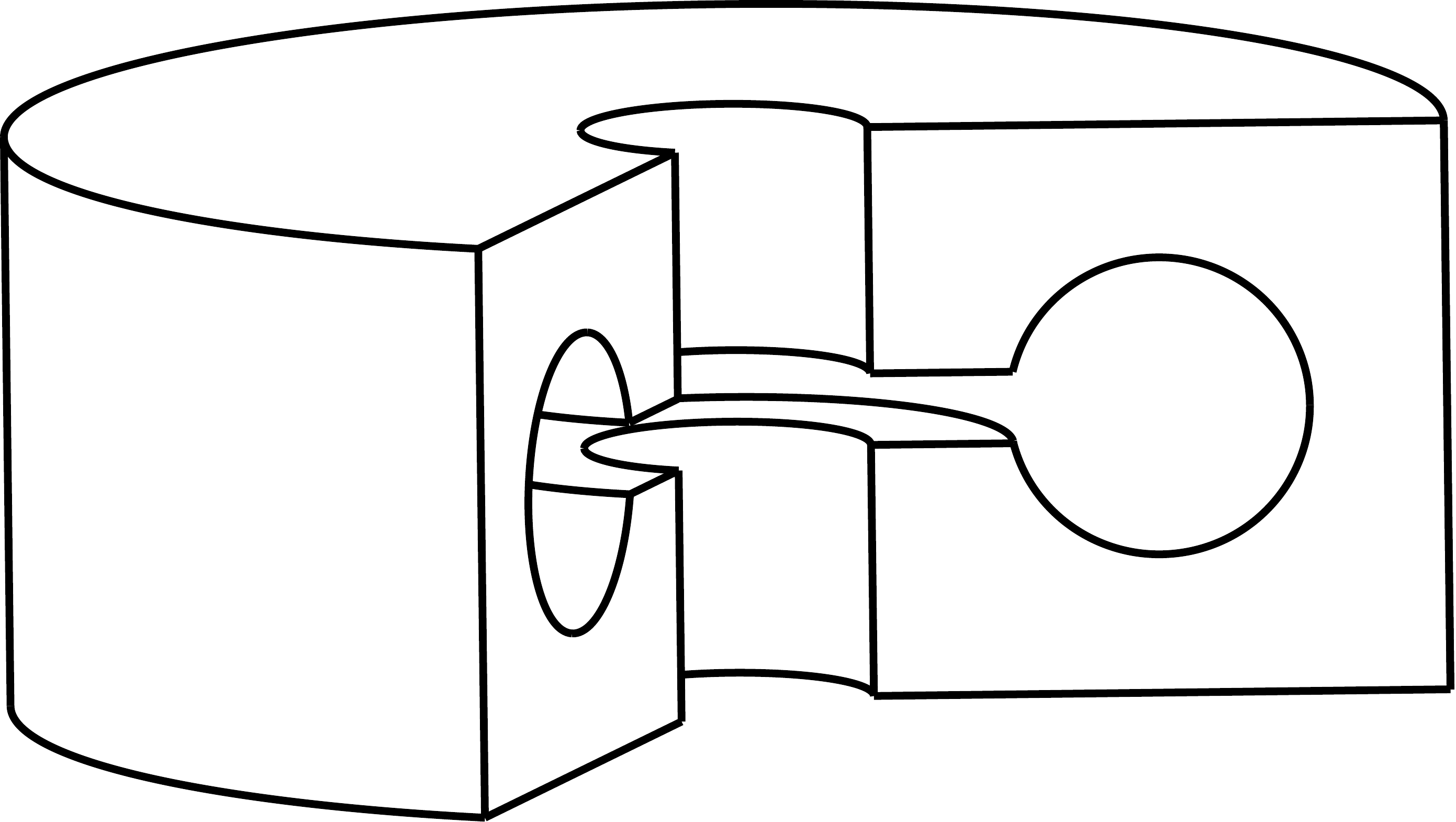}\\
(c)~\includegraphics[width=0.4\columnwidth]{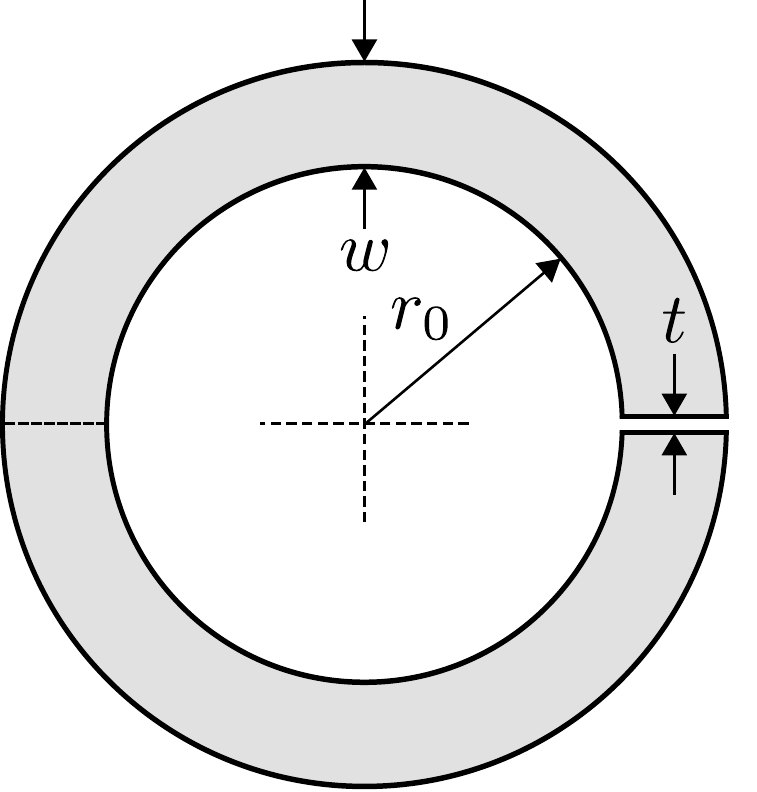} & \qquad(d)~\includegraphics[width=.75\columnwidth]{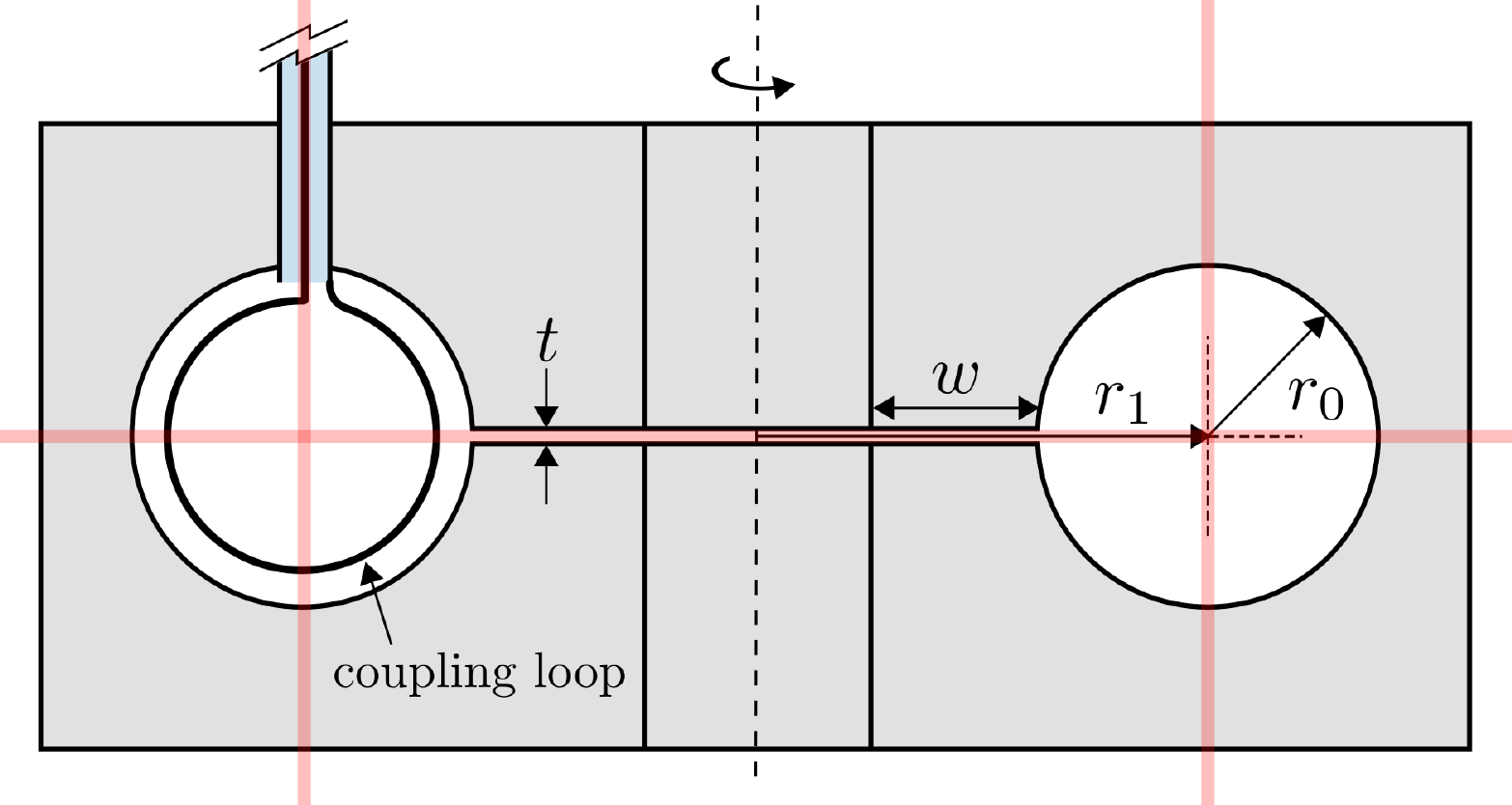}
\end{tabular}
}
\caption{Schematic drawings of the LGR designs. (a) A CLGR of length $\ell$.  The dashed line represents a conducting joint between two identical halves used to form the complete resonator.  (b) A TLGR with a section cut away to expose the interior.  (c) A cross-sectional view of the CLGR with the critical dimensions labeled.  (d) A cross-sectional view of the TLGR with its critical dimensions labeled.  A coupling loop is also shown suspended within the bore of the resonator.  The semi-transparent lines indicate the position of the coarse divider discussed in Section~\ref{sub:divide}.}
\label{fig:schematics}
\end{figure*}

Loop-gap resonators (LGRs) are electrically-small  high-Q resonators that can be used to make sensitive measurements of the electromagnetic (EM) properties of materials at RF and microwave frequencies~\cite{Hardy:1981, Froncisz:1982}.  Shown in Fig.~\ref{fig:schematics}(a), the cylindrical LGR (CLGR) consists of a conducting tube with a narrow slit cut along its length.  This structure can be accurately modeled as a series $LRC$ circuit.  The effective capacitance and inductance are determined by the geometry of the resonator gap and bore, respectively.  Currents run along the inner surface of the LGR bore and power dissipation associated with both conductive and radiative losses contribute to the resontor's effective resistance~\cite{Hardy:1981, Bobowski:2013}.  Radiative losses can be suppressed by joining the two ends of the CLGR to form a toroid that confines the magnetic fields~\cite{Bobowski:2016}.  A schematic of the toroidal LGR (TLGR) is shown in Fig.~\ref{fig:schematics}(b).  Parts (c) and (d) of Fig.~\ref{fig:schematics} show cross-sectional views of the CLGR and TLGR with the important dimensions labeled.  Part (d) of the figure also shows a coupling loop suspended within the bore of the TLGR.  The coupling loop is made by short-circuiting the center conductor of a coaxial cable to the outer conductor and is used to inductively couple signals into and out of the resonator.  In the same way, CLGRs can be excited and probed by placing coupling loops near the ends of the resonator bore~\cite{Rinard:1993}.       

\section{Probing EM material properties}

In this section, we give a simple description of how LGRs can be used to make sensitive measurements of the complex permittivity and conductivity of materials. The resonant frequency $\omega_0$ and the resonator quality factory $Q_0$ are modified by the electrical properties of materials placed in the LGR gap.  Current flow across the LGR gap has two components: 1) a displacement current that depends on the permittivity of the material and 2) charge transfer via a conduction current that depends on the conductivity of the material.

The gap admittance  is given by $Y_g=j\omega\varepsilon_r C_0+R_x^{-1}$, where $\omega$ is angular frequency, \mbox{$\varepsilon_r=\varepsilon^\prime-j\varepsilon^{\prime\prime}$} is the complex permittivity of the gap material, $C_0$ is the capacitance of the gap when it is empty, and $R_x$ is the resistance due to the gap geometry and the conductivity $\sigma$ of the gap material.  Expressing $R_x^{-1}=\sigma A/t=\sigma C_0/\varepsilon_0$, where $A$ is the gap area and $\varepsilon_0$ is the permittivity of free space, allows one to write
\begin{equation}
    Y_g=\omega\left[j\varepsilon^\prime + \left(\varepsilon^{\prime\prime}+\frac{\sigma}{\omega\varepsilon_0}\right)\right]C_0.
\end{equation}

Assuming radiative losses have been suppressed using either a CLGR with an EM shield or a TLGR, the effective resistance of the LGR can be written as $R=R_0\sqrt{\omega/\omega_0}$, where $R_0$ is the resistance at the resonant frequency $\omega_0$ of the empty resonator and the $\omega^{1/2}$ frequency dependence is due to the skin depth $\delta$.   In this case, the impedance of the LGR with a filled gap becomes $Z=R_1 + j\omega L_0 +\left(j\omega C_1\right)^{-1}$ where
\begin{align}
    R_1 &=R_0\sqrt{\frac{\omega}{\omega_0}}+\frac{1}{\omega C_0}\left[\frac{\varepsilon^{\prime\prime}+\sigma/\left(\omega\varepsilon_0\right)}{\left(\varepsilon^\prime\right)^2+\left(\varepsilon^{\prime\prime} +\sigma/\left(\omega\varepsilon_0\right)\right)^2}\right]\\
    C_1 &=C_0\left[\frac{\left(\varepsilon^\prime\right)^2+\left(\varepsilon^{\prime\prime} +\sigma/\left(\omega\varepsilon_0\right)\right)^2}{\varepsilon^\prime}\right],
\end{align}
and $\omega_0=1/\sqrt{L_0C_0}$ is the resonant frequency of the empty resonator.

The resonant frequency $\omega_1$ and quality factor $Q_1$ of the gap-filled resonator at $\omega=\omega_1$ can be calculated using \mbox{$\omega_1=1/\sqrt{L_0C_1}$} and $Q_1^{-1}=R_1\sqrt{C_1/L_0}$.  The results are
\begin{align}
    \left(\frac{\omega_1}{\omega_0}\right)^2 &=\frac{\varepsilon^\prime}{\left(\varepsilon^\prime\right)^2+\left(\varepsilon^{\prime\prime}+\sigma/\left(\omega_1\varepsilon_0\right)\right)^2}\label{eq:w1}\\
    \frac{1}{Q_1} &=\frac{1}{Q_0}\sqrt{\frac{\omega_0}{\omega_1}}+\frac{\varepsilon^{\prime\prime}+\sigma/\left(\omega_1\varepsilon_0\right)}{\varepsilon^\prime}.\label{eq:Q1}
\end{align}
Therefore, measurements of $\omega_0$ and $Q_0$ with the LGR gap empty, followed by measurements of $\omega_1$ and $Q_1$ with the gap filled, can be combined with (\ref{eq:w1}) and (\ref{eq:Q1}) to determine $\varepsilon^\prime$ and $\varepsilon^{\prime\prime}+\sigma/\left(\omega_1\varepsilon_0\right)$.  If either of the $\varepsilon^{\prime\prime}$ or $\sigma$ loss terms are dominant, as is often the case, then either $\varepsilon^\prime$ and $\varepsilon^{\prime\prime}$ or $\varepsilon^\prime$ and $\sigma$ can be independently determined.  This type of analysis has been used to study the EM properties of air, water, saltwater, liquid nitrogen, and methyl alcohol~\cite{Bobowski:2013, Bobowski:2017}.

It is also worth noting that, in an analogous way, the magnetic properties of materials can be investigated by filling the bore of a LGR with the material of interest~\cite{Bobowski:2015, Bonn:1991, Hardy:1993, Dubreuil:2019, Bobowski:2018, Madsen:2020}.

\section{Inductive Power Transfer}
\begin{figure*}[t]
\centerline{(a)~\includegraphics[height=5.4cm]{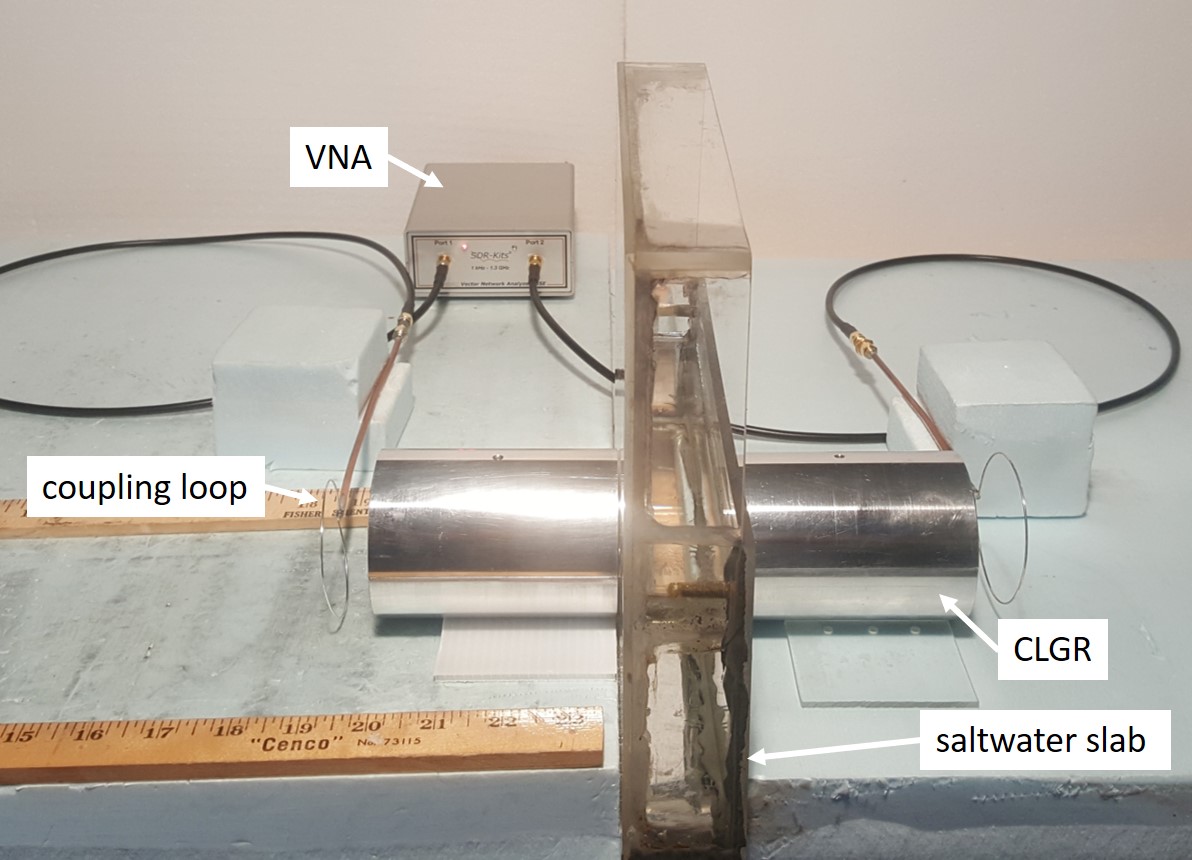} \quad (b)~\includegraphics[height=5.4cm]{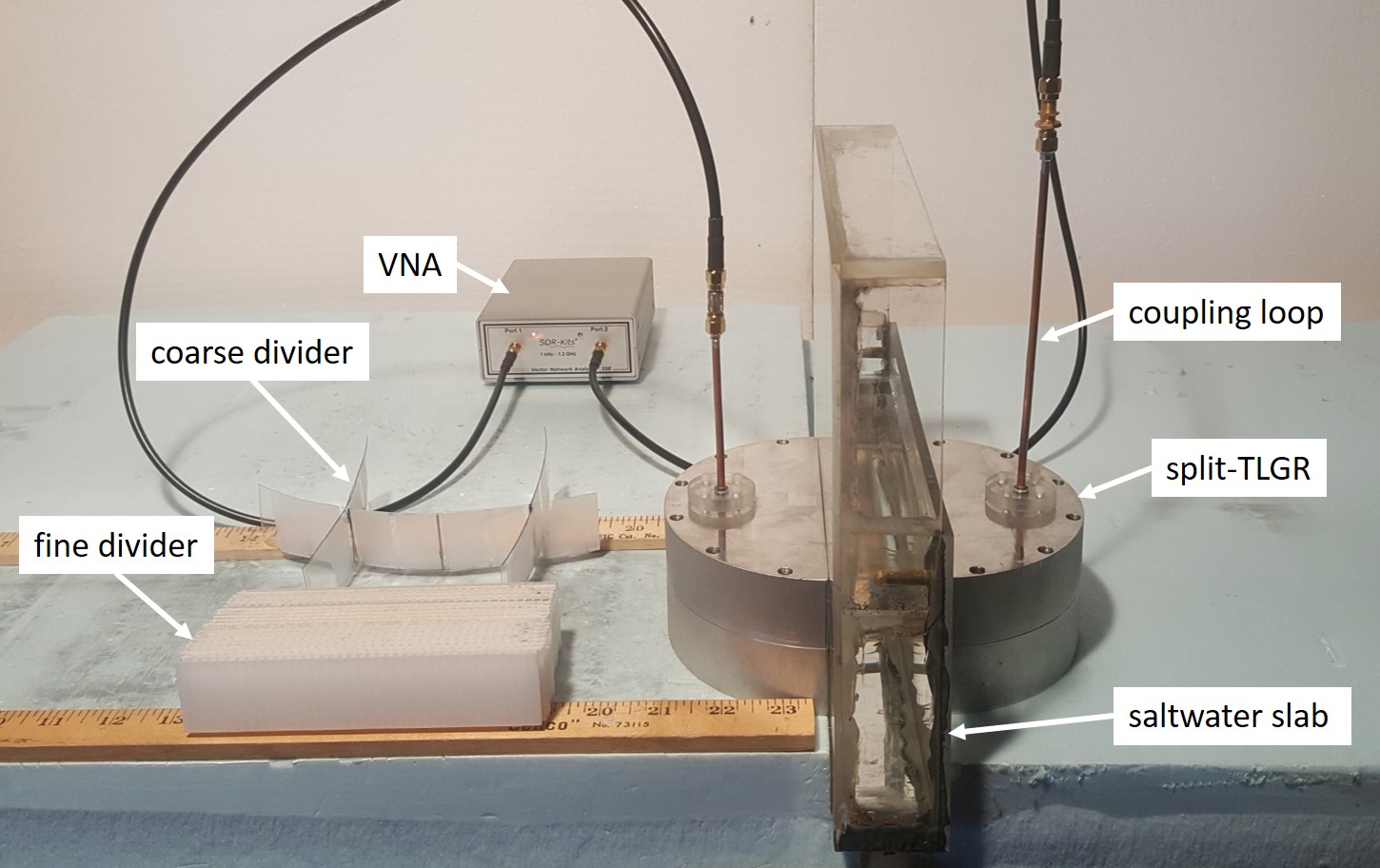}}
\caption{Photographs of IPT through a \SI[number-unit-product={\text{-}}]{3.7}{\centi\meter} thick slab of saltwater for both the (a) CLGR and (b) split-TLGR systems. The coarse and fine dividers used with the split-TLGR system are also shown in (b).  The \SI{102}{\micro\meter} gaps in the CLGRs are so narrow that they are not clearly visibile in the photograph.}
\label{fig:expt}
\end{figure*}

More recently, we have developed efficient mid-range inductive power transfer (IPT) systems using LGR transmitters and receivers~\cite{Roberts:2020}.  The term {\it mid-range} implies wireless energy exchange over distances that are several times the largest dimension of the transmitter/receiver~\cite{Soljacic:2007, Karalis:2008}.  Figures~\ref{fig:expt}(a) and (b) show the experimental setups of the CLGR and TLGR IPT systems, respectively.  In these figures, power is transferred wirelessly through a slab of saltwater.  
In the case of the TLGR system, the transmit and receive resonators are formed by dividing a complete TLGR with resonant frequency $\omega_0$ into two equal halves.  This division of the of the TLGR does not significantly alter the flow or distribution of charge in the structures and results in a pair of identical split-TLGRs with resonant frequencies approximately equal to $\omega_0$.  

Compared to helical and spiral resonators typically used in IPT systems, LGRs have the advantage that they allow for some shaping of the EM fields in the space surrounding the power-transfer link.  Specifically, electric fields are strongly confined to the narrow gap of the LGRs.  For IPT through a conductive medium, such as saltwater, this is an important advantage because excluding the conducting medium from the gap region effectively eliminates electric field power dissipation which is proportional to $\sigma E^2$.  Filling the gap with a low-loss dielectric, such as Teflon or \ce{Al2O3}, is a simple way to isolate the electric fields in the gap from the conducting medium.  For IPT through a conductive medium, there will also be power losses associated with the oscillating magnetic fields.  We consider this source of dissipation in Section~\ref{sec:dissipation}. 

The split-TLGR system has the additional advantage that the magnetic field strength outside the resonators is weak everywhere except between the transmitter and receiver~\cite{Roberts:2020}.  This field configuration limits the exposure of nearby individuals to EM fields, which is especially important in high-power applications.  For power transfer through a conducting medium, the CLGR configuration has magnetic power dissipation throughout the entire volume of space surrounding the pair of resonators.  In contrast, for the TLGR configuration, the magnetic field dissipation is primarily confined to the spatial volume directly between the bores of the transmit and receive resonators.

\section{Power Dissipation by a Conducting Medium}\label{sec:dissipation}
\begin{figure}[t]
\centerline{\includegraphics[bb = 56 584 363 736, clip = true, width=0.85\columnwidth]{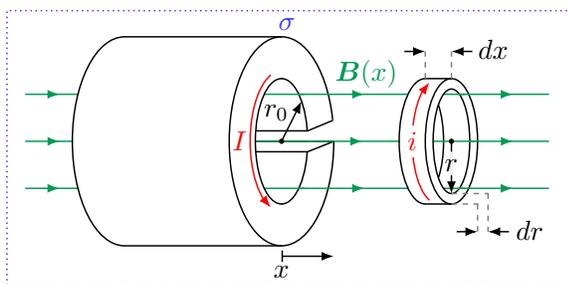}}
\caption{Schematic of the geometry used to estimate the magnetic power dissipated by the conducting medium.
}
\label{fig:geometry}
\end{figure}

We now present an approximate calculation of the power dissipation expected from an oscillating magnetic field in a conducting medium.  Although not rigorous, the results identify parameters that are important for practical designs.  Figure~\ref{fig:geometry} shows an example of a transmit or receive resonator  immersed in a medium of uniform conductivity $\sigma$.  Although the figure shows a CLGR, the calculations that follow can be applied to both the CLGR and split-TLGR geometries.

We assume an oscillating magnetic field of the form \mbox{$B(x)=B_0(x)\cos\omega_0 t$}, where $\omega_0$ is the resonant frequency of the LGR. The $x$-axis is chosen to coincide with the axis of the LGR bore.  In any plane perpendicular to the $x$-axis, $B_0$ is assumed to be uniform for $r\le r_0$ and zero for $r>r_0$.  Here, $r$ is the radial distance measured from the $x$-axis and $r_0$ is the radius of the LGR bore.  In the region between the transmit and receive resonators, $B_0$ will initially decrease as it moves away from the transmit resonator and then increase as it approaches the receive resonator.  The power dissipation is estimated by first finding the induced emf $\mathcal{E}$ around a circular loop of radius $r$ due to the changing magnetic flux.  We then estimate the resistance along the path followed by the resulting current.  Next, the power dissipation associated with each infinitesimal current loop is calculated.  Finally, the contributions from all current loops within a plane of width $dx$ are summed.  

Consider the infinitesimal ring of inner radius $r < r_0$, outer radius $r + \mathop{}\!d r$, and width $\mathop{}\!d x$ as shown in Fig.~\ref{fig:geometry}.  The magnetic flux $\Phi$ through the ring induces a current $i$ which flows along the circumference of the ring and through a cross-sectional area given by $\mathop{}\!d r \mathop{}\!d x$.  Therefore, the conductance of this infinitesimal ring is $\delta G=\sigma\,dr\,dx/\left(2\pi r\right)$. The induced emf $\mathcal{E}$ is calculated from $-d\Phi/dt$ which, in the assumed geometry, is approximately $\mathcal{E}\approx \pi r^2\omega_0 B_0(x)\sin\omega_0 t$. As a result, the power dissipation associated with the infinitesimal current loop of Fig.~\ref{fig:geometry} is given by
\begin{equation}
    \delta P=\mathcal{E}^2\delta G = \left(\pi/2\right) r^3\sigma\omega_0^2 B_0^2(x) \sin^2\omega_0 t\, dr\, dx.\label{eq:deltaP}
\end{equation}

The power dissipated by all current loops in a disk of thickness $dx$ is obtained by integrating $\delta P$ with respect to $r$ from zero to $r_0$, the radial range over which $B_0(x)$ is assumed to be non-zero and constant.  Evaluating this integral and taking a time average over one period yields
\begin{equation}
    \left\langle dP\right\rangle=\left\langle\int_{r=0}^{r_0}\delta P\right\rangle = \left(\pi/16\right) r_0^4\sigma\omega_0^2 B_0^2(x)\, dx.\label{eq:dP}
\end{equation}
Finding the total power dissipated in the space between the transmit and receive resonators requires an integration with respect to $x$ and a suitable model for the spatial dependence of $B_0(x)$. 

Fortunately, (\ref{eq:dP}) already reveals a number of useful insights.  First, the magnetic power dissipated by the conductive medium is proportional to the conductivity $\sigma$.  Second, for a fixed conductivity, power dissipation can be reduced by decreasing either $\omega_0$ or $r_0$. In Section~\ref{sec:expt}, we describe a set of experiments designed to separately test each one of these inferences.  We first note that since $\omega_0\propto 1/r_0$, the frequency can be decreased by increasing the size of the LGR~\cite{Hardy:1981, Bobowski:2013}.  However, this is not a good strategy because magnetic power loss varies by  $r_0^4$ which is much more significant than the $\omega_0^2$ dependence shown  in (\ref{eq:dP}).  Instead, it is best to design the LGR to be as small as possible and then lower the resonant frequency by filling the gap with a low-loss and high-permittivity dielectric.  A potential dielectric material for the capacitive gap is Mg-doped $\left(\mathrm{Ba},\mathrm{Sr}\right)\mathrm{TiO}_3$ which can simultaneously have very low loss tangents and relative permittivities as high as $10^3$~\cite{Nenasheva:2010, Song:2016}.

\begin{figure*}[t]
\centerline{
(a)\includegraphics[width=0.9\columnwidth]{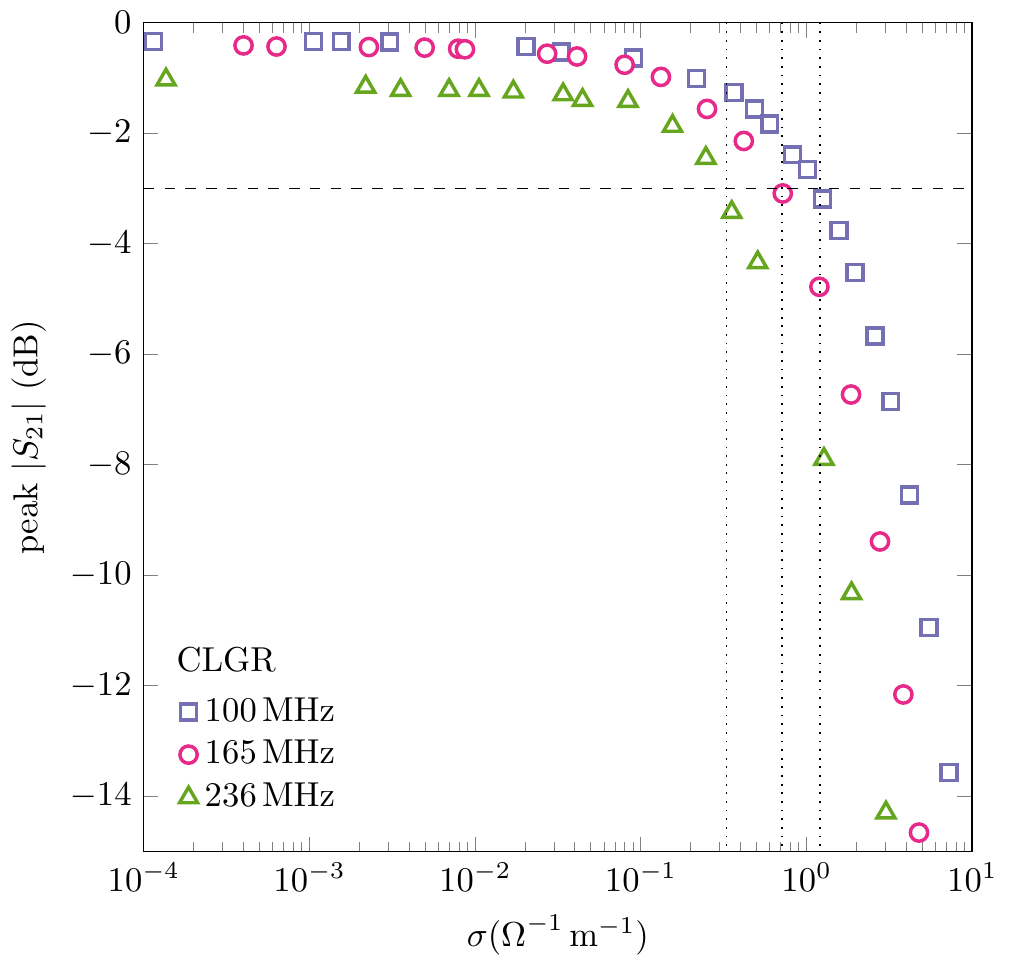} \qquad (b)\includegraphics[width=0.9\columnwidth]{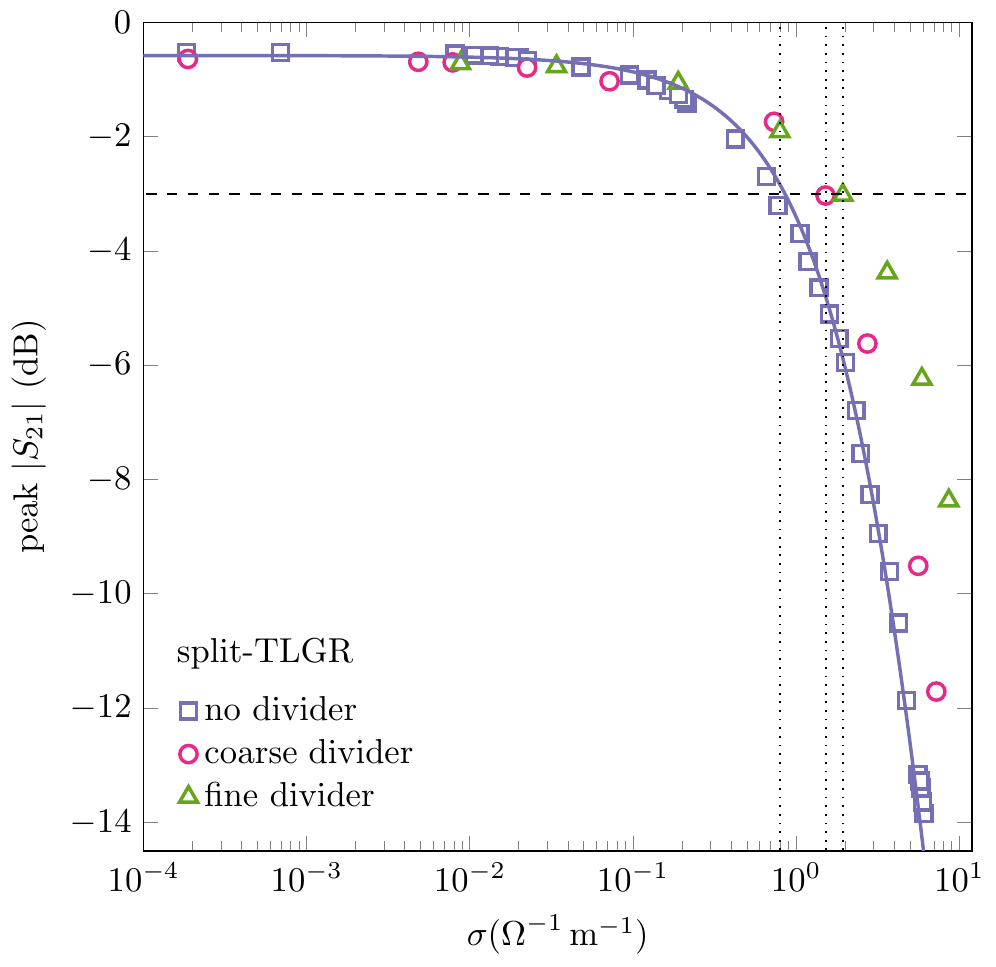}
}
\caption{Peak $\left\vert S_{21}\right\vert$ as a function of conductivity.  The horizontal dashed lines are placed at \SI{-3}{\decibel}.  The vertical dotted lines indicate the critical conductivity of the various systems investigated.  (a) CLGR IPT with the system resonant frequency tuned to three different values. The critical conductivity decreases with increasing frequency.  (b) TLGR IPT with and without dividers used to suppress induced current loops in the saltwater.  The solid line is a fit to the circuit model shown in Fig.~\ref{fig:circuit}.}
\label{fig:peakS21}
\end{figure*}

\section{Experiments and Results}\label{sec:expt}
The experimental setups for IPT through saltwater using CLGRs and split-TLGRs are shown in Figs.~\ref{fig:expt}(a) and (b), respectively.  All of the transmission coefficient ($S_{21}$) data reported in this paper were acquired using an SDR-Kits DG8SAQ vector network analyzer (VNA). A tank made from sheets of acrylic and epoxy was used to contain the saltwater.  The tank was \SI{44.0}{\milli\meter} thick and made using acrylic sheets that were \SI{3.7}{\milli\meter} thick.  The width and height of the tank were large enough that the magnetic field linking the transmit and receive resonators passed through the \SI[number-unit-product={\text{-}}]{36.6}{\milli\meter} thick saltwater slab. 

The design details of the LGRs are described elsewhere~\cite{Roberts:2020}.  However, we note that, with a Teflon dielectric filling the gaps, the resonant frequencies of the CLGRs and split-TLGRs were \SI{100}{\mega\hertz} and \SI{120}{\mega\hertz}, respectively. 

For both the CLGR and split-TLGR systems, data was collected as follows: First, the water tank was filled with \SI{1}{L} of deionized water with a base resistivity of \SI{15.4}{\mega\ohm\centi\meter}.  The LGR transmitter and receiver were then placed in contact with the two sides of the tank and the coupling loops were tuned to achieve optimal power transfer efficiency.  Next, the VNA was used to record an $\left\vert S_{21}\right\vert$ frequency sweep and the conductivity of the water was measured using a Beckman RC-16C conductivity bridge.  This procedure was repeated as \ce{NaCl} was added to the water in small amounts at a time.  For each concentration of \ce{NaCl}, the salt was allowed to completely dissolve and the tuning of the coupling loops was refined before acquiring the data.  

\begin{figure}[t]
\centerline{\includegraphics[width=0.92\columnwidth]{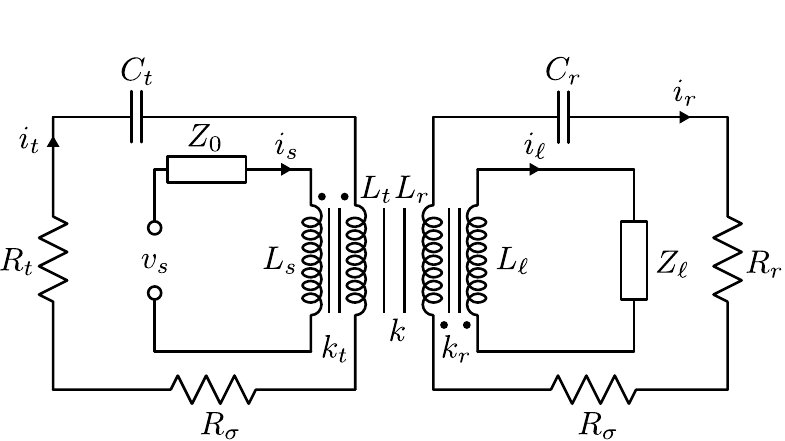}}
\caption{An equivalent circuit model of a four-coil IPT system operating in a conducting medium.  The signal source has output impedance $Z_0$ and supplies voltage $v_s$ and $Z_\ell$ is the load impedance.  The resistance $R_\sigma$ divides the induced emf between the transmitter/receiver resonators and the conducting medium.}
\label{fig:circuit}
\end{figure}
Figures~\ref{fig:peakS21}(a) and (b) show the peak value of $\left\vert S_{21}\right\vert$ in decibels as a function of conductivity for the CLGR and split-TLGR IPT systems, respectively.  For all of the various datasets shown, peak $\left\vert S_{21}\right\vert$ is approximately flat at low conductivity before dropping off steeply above a critical conductivity that we denote $\sigma_c$.  Qualitatively, this behaviour can be understood in terms of the equivalent circuit model of a four-coil IPT system shown in Fig.~\ref{fig:circuit}.  In this circuit, $R_t$ and $R_r$ represent the intrinsic effective resistances of the LGRs and $R_\sigma$ accounts for additional magnetic losses due to the saltwater.  At low conductivity, $R_t,\, R_r\gg R_\sigma$ and $\left\vert S_{21}\right\vert$ is approximately independent of $\sigma$.  However, above $\sigma_c$, $R_\sigma$ dominates and the $\left\vert S_{21}\right\vert$ peak drops as $\sigma$ is increased.  The solid line in Fig.~\ref{fig:peakS21}(b) is a fit to the split-TLGR data using a model for $\left\vert S_{21}\right\vert$ calculated from the equivalent circuit.  The fit assumed $R_\sigma = a \sigma$, where $a$ is the only free fit parameter and depends on the experimental geometry.  The other circuit parameters were obtained from a separate analysis of the split-TLGR system when transmitting power across an air gap~\cite{Roberts:2020}.  The fit is excellent and returned a best-fit value of $a=\SI[{scientific-notation = true, separate-uncertainty = true}]{5.81(7)e-2}{\ohm^2\meter}$. 

\subsection{Power transfer efficiency versus frequency}\label{sub:divide}
The CLGR and split-TLGRs were each made from two identical halves that bolt together to form the complete resonator.  The dashed line in Fig~\ref{fig:schematics}(a) indicates the conducting joint in the CLGR.   Inserting thin strips of copper tape, each \SI{89}{\micro\meter} thick,  between this joint provided a simple way of increasing the gap dimension $t$ (thereby decreasing the capacitance) while making only a negligible change to $r_0$ and the inductance.  This strategy, combined with extracting the Teflon dielectric from the gap, was implemented to increase the operating frequency of the CLGR IPT link above the base frequency of \SI{100}{\mega\hertz} while keeping the resonator size $2\left(r_0+w\right)$ fixed.  Figure~\ref{fig:peakS21}(a) shows the peak $\left\vert S_{21}\right\vert$ as a function of conductivity for three different CLGR resonant frequencies.  As seen from the frequency dependence of the critical conductivity $\sigma_c$, the system performance degrades as IPT operating frequency is increased.  This observation is consistent with (\ref{eq:dP}) and the analysis presented in Section~\ref{sec:dissipation}.   

\subsection{Suppressing large-radius current loops}
Finally, we repeated the peak $\left\vert S_{21}\right\vert$ versus $\sigma$ measurements for the split-TLGR system after inserting dividers into the saltwater bath.  The dividers were used to partition the saltwater in an effort to suppress large-radius current loops.  Equation (\ref{eq:dP}) suggests that the magnetic power dissipation is a strong function of $r_0$, the maximum radius of the induced current loops.  Therefore, using dividers to reduce the average size of the current loops is expected to enhance the IPT efficiency.

The semi-transparent lines in Fig.~\ref{fig:schematics}(d) indicate the position of the coarse divider used in our experiments.  The coarse divider, made from thin plastic strips, is also shown in Fig.~\ref{fig:expt}(b). The circular data points in Fig.~\ref{fig:peakS21}(b) show peak $\left\vert S_{21}\right\vert$ as a function of $\sigma$ when the coarse divider was in place.  As anticipated, the power transfer efficiency improved and $\sigma_c$ doubled from \SI{0.75}{\ohm\tothe{-1}\meter\tothe{-1}} without a divider to \SI{1.5}{\ohm\tothe{-1}\meter\tothe{-1}} with the coarse divider.  We repeated the measurements using a fine divider, also shown in Fig.~\ref{fig:expt}(b), made by stacking and gluing sheets of corrugated plastic with parallel $\SI{3.5}{\milli\meter}\times\SI{3.5}{\milli\meter}$ channels.  To get the saltwater to penetrate into the narrow channels, \SI{5}{\milli L} of dish soap was added to act as a surfactant.  We first verified that the added soap did not alter the $\left\vert S_{21}\right\vert$ versus $\sigma$ data with no divider in place.  As shown in Fig.~\ref{fig:peakS21}(b), the fine divider caused $\sigma_c$ to increase to \SI{2.0}{\ohm\tothe{-1}\meter\tothe{-1}}.  At a conductivity of \SI{5.0}{\ohm\tothe{-1}\meter\tothe{-1}}, typical of seawater, the coarse and fine dividers improved the peak power transfer efficiency by \num{3.4} and \SI{6.6}{\decibel}, respectively. 

\section{Conclusion}
We have demonstrated IPT through saltwater using LGR transmitters and receivers.  Below a critical conductivity $\sigma_c$, dissipation is dominated by losses that are intrinsic to the LGRs.  However, above $\sigma_c$, magnetic power dissipation by the saltwater becomes important and the power transfer efficiency rapidly drops.  We showed that the critical conductivity increases as the LGR resonant frequency is decreased for a resonator of fixed size.  Using dividers to partition the saltwater volume, we found that restricting the size of the induced current loops in the conducting medium provided another means to enhance the power transfer efficiency.  All of these observations are consistent with an approximate calculation of the expected magnetic power dissipation due to a conducting medium.  Our results suggest that the highest efficiency will be achieved by simultaneously minimizing the resonator size and resonant frequency.  For LGRs, these design criteria are best met by filling the narrow gap of the resonators with a high-$\varepsilon_r$ and low-loss dielectric.

We are currently experimenting with LGR transmitters and receivers equipped with watertight seals used to exclude saltwater from the gaps and bores of the resonators.  The seals allow us to completely submerge the LGRs in a saltwater bath and work with testbeds that more closely replicate the conditions expected in practical applications.

\begin{thebibliography}{00}
\bibitem{Hardy:1981}

W. N. Hardy and L. A. Whitehead, ``Split-ring resonator for use in magnetic resonance from 200--2000 {MHz},'' {\it Rev.\ Sci.\ Instrum.}, vol.~52, no.~2, pp.~213--216, Feb.~1981.

\bibitem{Froncisz:1982}
W. Froncisz and J. S. Hyde, ``The loop-gap resonator: {A} new microwave lumped circuit {ESR} sample structure,'' {\it J.\ Magn.\ Reson.}, vol.~47, no.~3, pp.~515--521, May~1982.

\bibitem{Bobowski:2013}
J. S. Bobowski, ``Using split-ring resonators to measure the electromagnetic properties of materials: An experiment for senior physics undergraduates,'' {\it Am.\ J.\ Phys.}, vol.~81, no.~12, pp.~899--906, Dec.~2013.

\bibitem{Bobowski:2016}
J. S. Bobowski and H. Nakahara, ``Design and characterization of a novel toroidal split-ring resonator,'' {\it Rev.\ Sci.\ Instrum.}, vol.~87, no.~2, pp.~024701, Feb.~2016.

\bibitem{Rinard:1993}
G. A. Rinard, R. W. Quine, S. S. Eaton, and G. R. Eaton, ``Microwave coupling structures for spectroscopy,'' {\it J.\ Magn.\ Reson.\ Ser.\ A}, vol.~105, no.~2, pp.~137--144, Nov.~1993.

\bibitem{Bobowski:2017}
J. S. Bobowski and A. P. Clements, ``Permittivity and conductivity measured using a novel toroidal split-ring resonator,'' {\it IEEE Trans.\ Microw.\ Theory Tech.}, vol.~65, no.~6, pp.~2132--2138, June~2017.

\bibitem{Bobowski:2015}
J. S. Bobowski, ``Using split-ring resonators to measure complex permittivity and permeability,'' in {\it Proc. Conf. Lab. Instruct. Beyond First Year College}, College Park, MD, USA, 215, pp. 20--23.

\bibitem{Bonn:1991}
D. A. Bonn, D. C. Morgan and W. N. Hardy, ``Split‐ring resonators for measuring microwave surface resistance of oxide superconductors,''  {\it Rev.\ Sci.\ Instrum.}, vol.~62, no.~7, pp.~1819--1823, July~1991.

\bibitem{Hardy:1993}
W. N. Hardy, D. A. Bonn, D. C. Morgan, R. Liang and K. Zhang, ``Precision measurements of the temperature dependence of \ensuremath{\lambda} in ${\mathrm{YBa}}_{2}$${\mathrm{Cu}}_{3}$${\mathrm{O}}_{6.95}$: Strong evidence for nodes in the gap function,'' {\it Phys. Rev. Lett.}, vol.~70, no.~25, pp.~3999--4002, June~1993.

\bibitem{Dubreuil:2019}
J. Dubreuil and J. S. Bobowski, ``Ferromagnetic resonance in the complex permeability of an {Fe$_3$O$_4$}-based ferrofluid at radio and microwave frequencies,'' {\it J.\ Magn.\ Magn.\ Mater.}, vol.~489, pp.~165387, Nov.~2019.

\bibitem{Bobowski:2018}
J. S. Bobowski, ``Probing split-ring resonator permeabilities with loop-gap resonators,'' {\it Can.\ J.\ Phys.}, vol.~96, no.~8, pp.~878--886, Aug.~2018.

\bibitem{Madsen:2020}
S. L. Madsen and J. S. Bobowski, ``The complex permeability of split-ring resonator arrays measured at microwave frequencies," {\it IEEE Trans.\ Microw.\ Theory Tech.}, vol.~86, no.~8, pp.~3547--3557, Aug.~2020.

\bibitem{Roberts:2020} 
D. M. Roberts, A. P. Clements, R. McDonald, J. S. Bobowski, and T. Johnson, ``Mid-range wireless power transfer at \SI{100}{\mega\hertz} using magnetically-coupled loop-gap resonators,'' 2021. [Online]. Available: arXiv:2103.14798.

\bibitem{Soljacic:2007}
A. Kurs, A. Karalis, R. Moffatt, J. D. Joannopoulos, P. Fisher and M. Solja\v{c}i\'{c}, ``Wireless power transfer via strongly coupled magnetic resonances,'' {\it Science}, vol.~317, no.~5834, pp.~83--86, Jul.~2007.

\bibitem{Karalis:2008}
A. Karalis, J. D. Joannopoulos and M. Solja\v{c}i\'{c}, ``Efficient wireless non-radiative mid-range energy transfer,'' {\it Ann. Phys.}, vol.~323, no.~1, pp.~34--48, Jan.~2008.

\bibitem{Nenasheva:2010}
E. A. Nenasheva, N. F. Kartenko, I. M. Gaidamaka, O. N. Trubitsyna, S. S. Redozubov, A. I. Dedyk and A. D. Kanareykin, ``Low loss microwave ferroelectric ceramics for high power tunable devices,'' {\it J.\ Eur.\ Ceram.\ Soc.}, vol.~30, no.~2, pp.~395--400, Jan.~2010.

\bibitem{Song:2016}
M. Song, P. Belov and P. Kapitanova, ``Wireless power transfer based on dielectric resonators with colossal permittivity,'' {\it Appl.\ Phys.\ Lett.}, vol.~109, no.~22, p.~223902, Dec.~2016.

\end{thebibliography}
\end{document}